\documentclass{aa}  
\usepackage{graphicx,natbib}
\bibpunct{(}{)}{;}{a}{}{,} 
\usepackage{txfonts}

\newcommand{\cpks}{ct~ks$^{-1}$}
\newcommand{\kms}{km~s$^{-1}$}

\newcommand{\ergps}{ergs~s$^{-1}$}

\newcommand{\ergpshz}{ergs~s$^{-1}$~Hz$^{-1}$}

\newcommand{\cxc}{\textit{Chandra}}

\begin{document}
   \title{A \cxc\  X-ray detection of the L dwarf binary Kelu-1}
   
   \subtitle{Simultaneous \cxc\  and Very Large Array observations}

   \author{M. Audard\inst{1,2}
          \and
	  R.~A. Osten\inst{3}
	  \fnmsep\thanks{Hubble Fellow}
	  \and
	  A. Brown\inst{4}
	  \and
	  K.~R. Briggs\inst{5}
	  \and
	  M. G\"udel\inst{5}
	  \and
	  E. Hodges-Kluck\inst{3,4}
	  \and
	  J.~E. Gizis\inst{6}
          }

   \offprints{\email{Marc.Audard@obs.unige.ch}}
   
   \institute{ISDC, Ch. d'Ecogia 16, CH-1290 Versoix, Switzerland
         \and
	  Observatoire de Gen\`eve, University of Geneva, Ch. des Maillettes 51, 1290 Sauverny, 
Switzerland
	\and
             Astronomy Department, University of Maryland, College Park, MD, USA
	\and
             Center for Astrophysics and Space Astronomy, University of Colorado,
	     Boulder, CO 80309-0389, USA
	\and
	     Paul Scherrer Institut, Villigen \& W\"urenlingen, 5232 Villigen PSI,
	     Switzerland
	 \and
		Department of Physics and Astronomy, 223 Sharp Lab, University of
		Delaware, Newark, DE 19716, USA
             }

   \date{Received June 15, 2007; accepted July 6, 2007}

 
  \abstract
   {Magnetic activity in ultracool dwarfs, as measured in X-rays and H$\alpha$, shows a steep decline after spectral 
   type M7-M8. So far, no L dwarf has been detected in X-rays. In contrast, L dwarfs may have higher radio activity than M dwarfs.}
   {We observe L and T dwarfs simultaneously in X-rays and radio to determine their level of magnetic activity in the context of the
   general decline of magnetic activity with cooler effective temperatures.}
   {The field L dwarf binary Kelu-1 was observed simultaneously with \cxc\  and the Very Large Array.}
   {Kelu-1AB was detected in X-rays with $L_{\rm X} = 2.9_{-1.3}^{+1.8} \times 10^{25}$~\ergps, while it remained undetected in the radio down to 
   a $3 \sigma$ limit of
   $L_{\rm R} \leq 1.4 \times 10^{13}$~\ergpshz. We argue that, whereas the X-ray and H$\alpha$ emissions decline in ultracool dwarfs with decreasing effective
   temperature, the radio luminosity stays (more or less) constant across M and early-L dwarfs. The radio surface
   flux or the luminosity may better trace magnetic activity in ultracool dwarfs than the ratio of the luminosity to the bolometric luminosity.
   }
   {Deeper radio observations (and at short frequencies) are required to determine if and when the cut-off in radio activity occurs in L and T dwarfs, and what kind
   of emission mechanism takes place in ultracool dwarfs.}

   \keywords{radio continuum: stars --
   		stars: activity --
		stars: coronae --
		stars: individual (Kelu-1) --
		stars: low-mass, brown dwarfs --
		X-rays: stars
               }

   \maketitle
%

\section{Introduction}
\label{sect:intro}

There is significant evidence that magnetic activity, commonly seen in low-mass stars, survives in ultracool dwarfs
\citep[e.g.,][]{tagliaferri90,fleming93,drake96,fleming00,rutledge00, 
martin02,schmitt02,fleming03,briggs04,stelzer04,berger05,hallinan06,osten06,stelzer06a,stelzer06b,phanbao07}.
Since the latter are fully convective, a different kind of magnetic dynamo mechanism than in the Sun
and in stars with tachoclines must take place in ultracool dwarfs, possibly due to turbulent magnetic fields. 

A common indicator of magnetic activity in late-type stars, X-rays have so far never been detected in stars later than
spectral type M9. \citet{stelzer03}, \citet{berger05}, and \citet{stelzer06b} reported the non-detection in X-rays of L dwarfs,
down to $<6.6\times 10^{24}$~\ergps. 
Studies of the X-ray emission in M dwarfs show a decline in emission after spectral type M7-M8 
\citep[e.g.,][]{fleming03,stelzer06a}, in parallel with
the decline in the chromospheric H$\alpha$ emission \citep{gizis00,mohanty03,west04}. 
The radio luminosity in late-type stars correlates over several decades with the X-ray luminosity \citep{guedel93a,benz94};
however, \citet{berger02} argued that ultracool dwarfs do not follow this correlation and suggested
an increase in radio activity
with cooler effective temperatures. Such an increase was further supported by \citet{berger05} and \citet{burgasser05} 
for late-M and early-L dwarfs, despite very low radio detection rates \citep[e.g.,][]{berger06}. 
The decline in H$\alpha$ and X-ray activity, and
the non-detections of late-L and T dwarfs in the radio, may be related to the highly neutral atmospheres of these dwarfs,
which decouples the magnetic fields from the photospheric gas \citep{meyer99,mohanty02} and could lead to unfavorable conditions for magnetic 
activity.

In this Letter, we present the results of simultaneous observations of the early-L field brown dwarf binary Kelu-1
\citep{ruiz97,liu05,gelino06} with \cxc\  and the Very Large Array (VLA). 
As mentioned earlier, no L dwarf has yet been detected in X-rays; this Letter presents,
therefore, the first X-ray detection of an L dwarf, while the dwarf remains undetected in the radio.

\section{The Kelu-1AB dwarf binary}
\label{sect:kelu1}

At a distance of $18.7 \pm 0.7$~pc, Kelu-1 was  found thanks to its high proper motion ($\mu \sim 0\farcs 3$/yr; \citealt{ruiz97,dahn02,scholz02,lodieu05}).
Its optical spectrum shows weak Li~{\sc i} absorption and H$\alpha$ in emission \citep{ruiz97,kirkpatrick99}.
Its age is difficult to assess, but it probably lies in the range $0.3-0.8$~Gyr \citep{liu05}. \citet{basri00} and \cite{mohanty03} measured a high rotation rate 
($v \sin i \sim 60$~\kms), 
supported by a 1.8~h photometric period in H$\alpha$ \citep{clarke02,clarke03}, which cannot be due to the orbit of a binary \citep{gelino06}. 
Kelu-1 remained undetected in the X-rays down to $L_{\rm X} < 7.3 \times 10^{27}$~\ergps\ \citep{neuhaeuser99},
 and in the radio ($S_{3.6} < 28~\mu$Jy, $3 \sigma$ limit; \citealt{krishna99}).

It was originally classified as an L2 dwarf; however, high spatial resolution images have
recently revealed its binary nature \citep{liu05,gelino06}. The binary is separated by about $0\farcs 3$ with a position angle of $\sim 220\degr$
\citep{liu05}, and there is evidence of orbital motion \citep{gelino06}. The latter authors estimate that Kelu-1A has spectral type L2 $\pm 1$, 
whereas Kelu-1B is slightly colder with a spectral type of L3.5 $\pm 1$, in line with the estimates (L1.5-L3 and L3-L4.5) of \citet{liu05} based on different methods.
\citet{gelino06} also give masses of $0.060 \pm 0.01$ and $0.055\pm 0.01~M_\odot$, bolometric luminosities
$\log (L_{\rm bol}/L_\odot) = -3.83$ and $-3.99$, and effective temperatures in the range $1900-2100$~K and $1700-1900$~K for Kelu-1A and B, respectively.

\section{Observations and data reduction}
\label{sect:data}

\begin{table}
\caption{Observation log for VLA  and \cxc\label{tab:log}}
\begin{tabular}{l c}        
\hline\hline                 
 \noalign{\vskip .8ex}%
\multicolumn{2}{c}{Position of Kelu-1AB (Equinox: J2000; Epoch:  J2006.41)}\\
 \noalign{\vskip .5ex}%
\hline
 \noalign{\vskip .5ex}%
 \multicolumn{2}{c}{Based on \citet{lodieu05}}\\
Right ascension\dotfill                 &       $13\fh 05\fm 40\fs 00$\\
Declination\dotfill                     &       $-25\fdg 41\farcm 06\farcs 1$\\
 \multicolumn{2}{c}{\cxc\ mean position}\\
Right ascension\dotfill                 &       $13\fh 05\fm 40\fs 01 \pm 0\fs 01$\\
Declination\dotfill                     &       $-25\fdg 41\farcm 05\farcs 9 \pm 0\farcs 10$\\
 \noalign{\vskip .5ex}%
 \hline
 \noalign{\vskip .5ex}%
 \multicolumn{2}{c}{VLA (Program S6570)}\\
 \noalign{\vskip .5ex}%
 \hline
 \noalign{\vskip .5ex}%
Observation date\dotfill                &       2006 May 31 0h36 -- 7h36 UT\\
Configuration \& Wavelength\dotfill     &       BnA \& 3.6 cm (X band)\\
Flux and phase calibrators\dotfill      &       1331$+$305 (3C 286) / 1258$-$223\\
 \noalign{\vskip .5ex}%
 \hline
 \noalign{\vskip .5ex}%
 \multicolumn{2}{c}{\cxc\  (ObsId 5421)}\\
 \noalign{\vskip .5ex}%
 \hline
 \noalign{\vskip .5ex}%
Observation date\dotfill                &       2006 May 31 0h04 -- 7h18 UT\\
Instrument \& Mode \dotfill             &       ACIS-S \& VFAINT\\
Total exposure\dotfill                  &       23.8~ksec\\
\hline
\end{tabular}
\end{table}

The \cxc\  observation was coordinated with the VLA. 
Table~\ref{tab:log} provides the details of the observations. 
The X-ray data were processed with the CIAO 3.3.0.1 software together with CALDB 3.2.2.
The VLA data were calibrated with the AIPS software.
A typical dwell cycle spent 2 minutes
on the phase calibrator and 9.5 minutes on Kelu-1. The total VLA on-source time in the  3.6~cm band was
4.4 hours.

   \begin{figure}[!t]
   \centering
   \includegraphics[width=.9\linewidth]{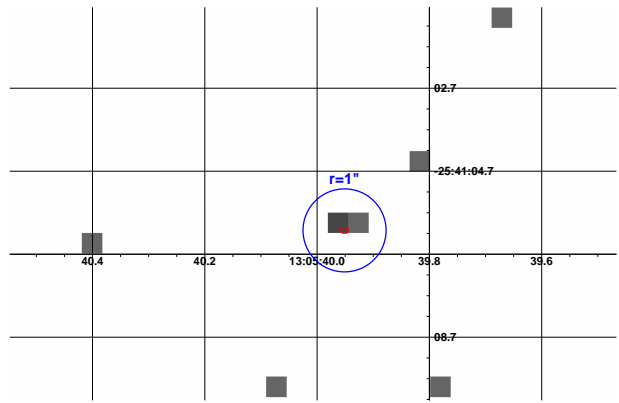}
      \caption{\cxc\ $0.2-6.0$~keV image centered on Kelu-1AB. The small red ellipse corresponds
      to the expected position (with errors) of Kelu-1. The extraction region is shown as a blue circle
      of 1\arcsec\  radius. The nearby background is very low, with sky pixels containing either 0 or 1 event.
	      }
	 \label{fig:cxcim}
   \end{figure}
   \begin{figure}[!t]
   \centering
   \includegraphics[width=.9\linewidth]{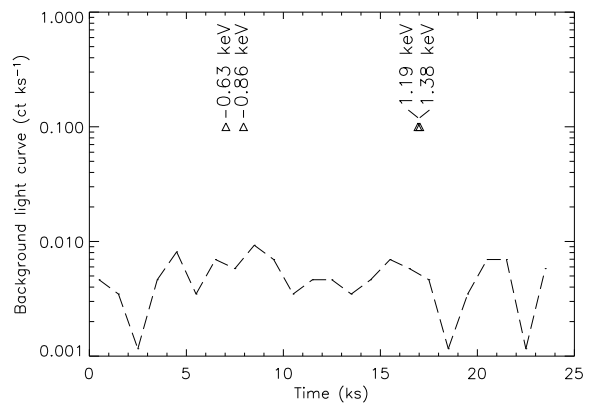}
      \caption{\cxc\ $0.2-6.0$~keV light curve of the background around Kelu-1AB, scaled down to the extraction area of the source.
      The 4 events at the position of Kelu-1AB are placed arbitrarily at a $y$-value of 0.1; the event energies
      are also labeled.
      }
         \label{fig:cxclc}
   \end{figure}

We used the proper motion properties of Kelu-1 as given by \citet{lodieu05} to determine the position of the binary at the epoch of
observations  (see Table~\ref{tab:log}).
We extracted X-ray events inside a circle centered at the position of Kelu-1 and with a radius of 1\arcsec\ 
(Fig.~\ref{fig:cxcim}). 
We used the energy range of $0.2-6.0$~keV to  reduce the background contamination. We obtained 4 events of energy $0.63$, $0.86$,
$1.19$ and $1.38$~keV.  There is no strong clustering of the event arrival times (the 3rd and 4th events are separated by
91.1~s, much longer than the frame time, 3.1~s), suggesting that the events do not arise from flares.
We estimated the background level by extracting events in a concentric annulus of 
 3\arcsec\  and 30\arcsec\ radii that avoided nearby sources. We extracted 102 background
events, which corresponds to a background level in the extraction region for Kelu-1 of less than 1 event (0.11 event). The average background
count rate was, thus, $0.005$~\cpks\  (Fig.~\ref{fig:cxclc}).
Using the \citet{kraft91} approach, the 68\% Bayesian confidence range for the number of source events for Kelu-1 is $2.18-6.27$.  
With the approach of \citet{ayres04}, we determined
that the X-ray detection of Kelu-1 was a $4.3 \sigma$ detection for a one-sided Gaussian distribution.

In a similar fashion as for $\epsilon$ Ind Bab \citep{audard05}, we simulated a single temperature plasma model (APEC 1.3.1;
\citealt{smith01}) with solar abundances
 in XSPEC to determine the conversion factor from count rate to X-ray luminosity. Using the distance of Kelu-1,
we estimate a 0.1-10~keV X-ray luminosity of $L_{\rm X} = 2.9 \times 10^{25}$~\ergps, 
while the 68\% Bayesian confidence range based on the \citet{kraft91} approach corresponds to $(1.6 - 4.7) \times 10^{25}$~\ergps.
Note that the above estimates work for plasma temperatures of 0.4 to 1 keV, which could be expected from old brown dwarfs. As mentioned in \citet{audard05}, the values can be higher by factors of
$1.2-2.0$ and even $3.25$ in the extreme case of a plasma temperature of $0.1$~keV. 

 In the radio regime, Kelu-1 remained undetected during the observation down to a rms flux density of 14 $\mu$Jy, i.e., 
a $3 \sigma$ upper limit for the radio luminosity at 3.6~cm of $L_{\rm R} < 1.76 \times 10^{13}$~\ergpshz. 
After combining with the archival VLA data from \citet{krishna99}, Kelu-1 still remains undetected down to an rms flux
density of 11 $\mu$Jy, i.e., $L_{\rm R} < 1.38 \times 10^{13}$~\ergpshz\ 
($3\sigma$).

\section{Discussion}
\label{sect:discussion}

   \begin{figure*}[!t]
   \centering
   \hfill\includegraphics[width=.45\linewidth]{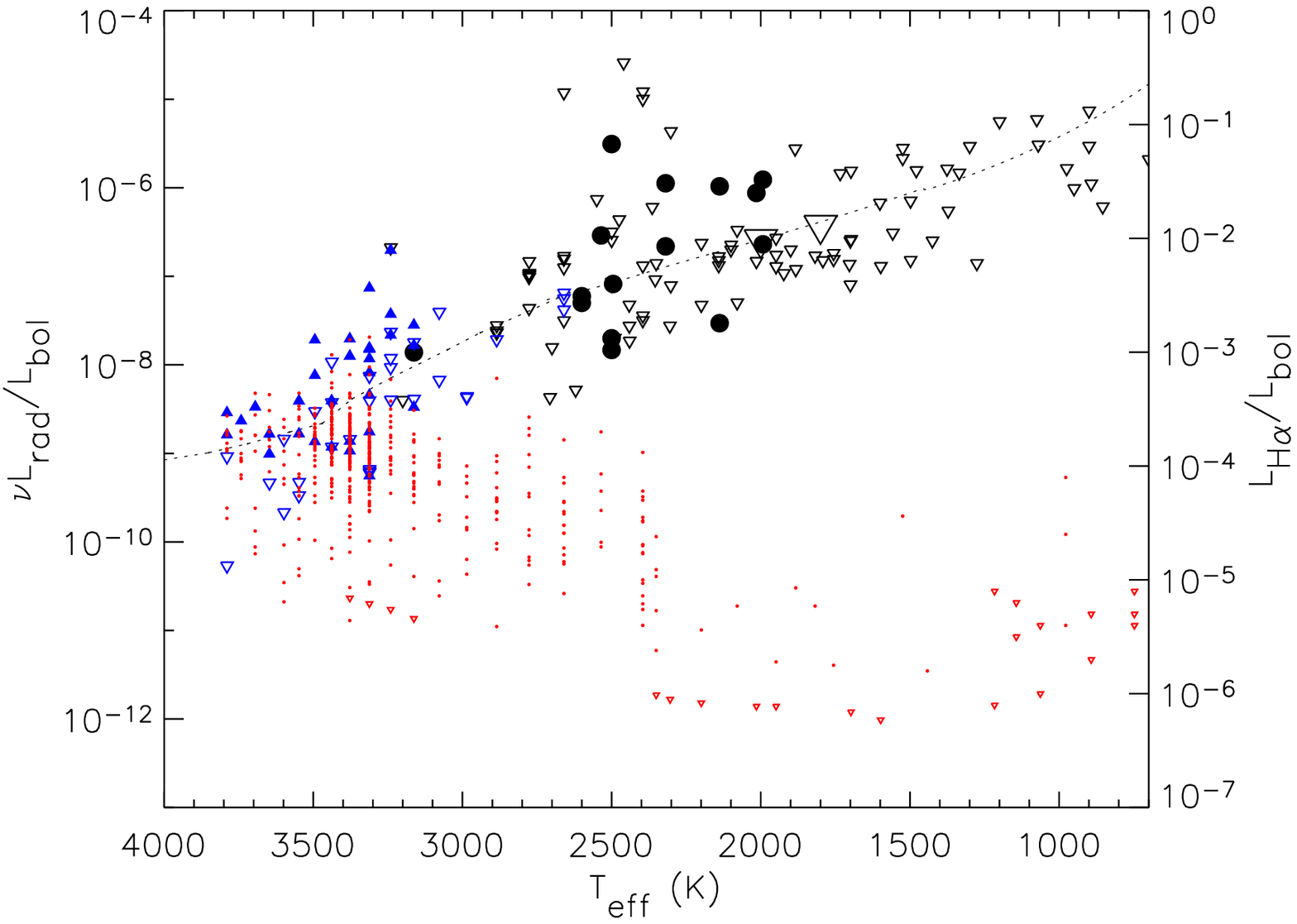}\hfill
   \includegraphics[width=.45\linewidth]{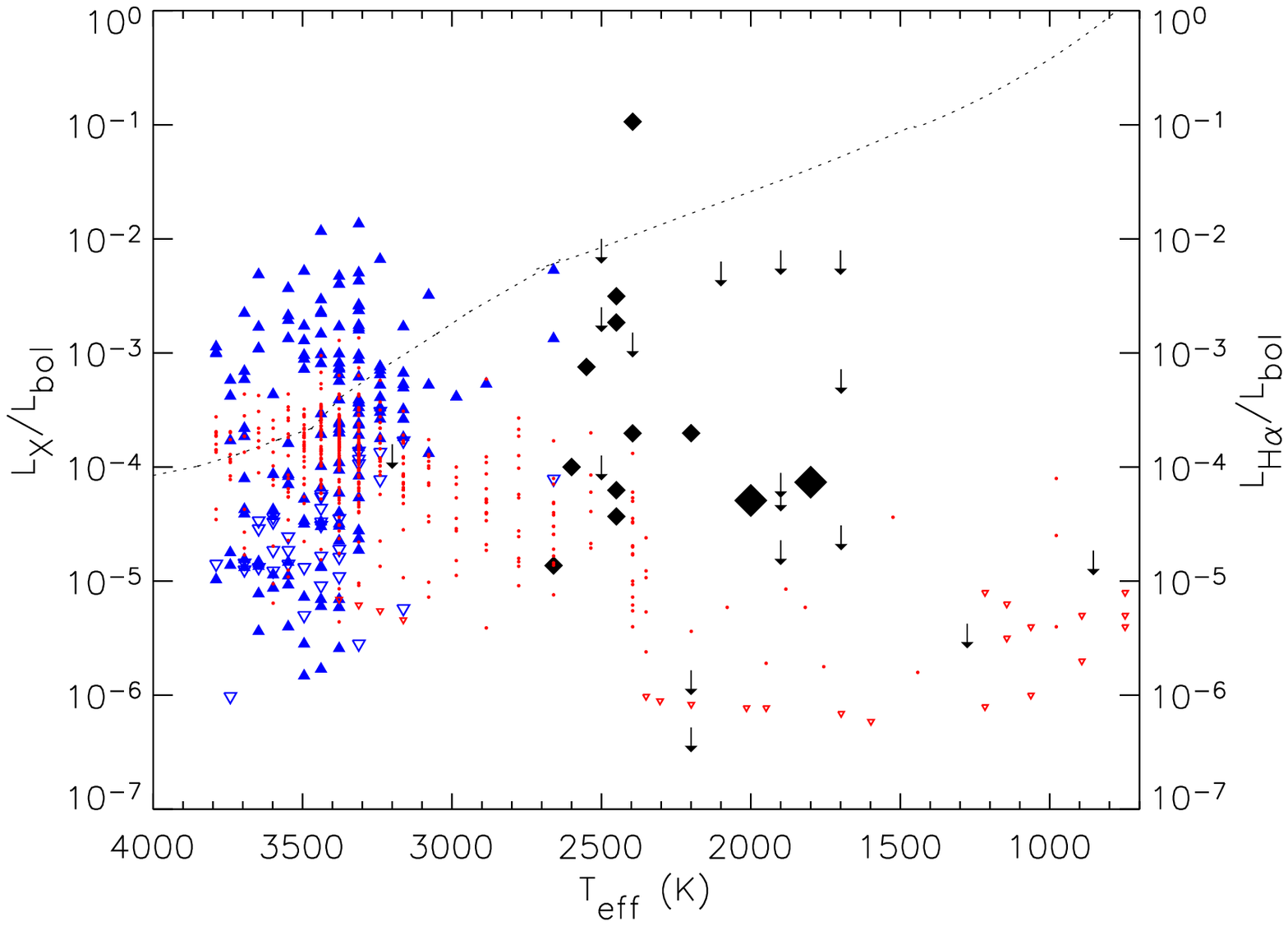}\\
   \hfill\includegraphics[width=.45\linewidth]{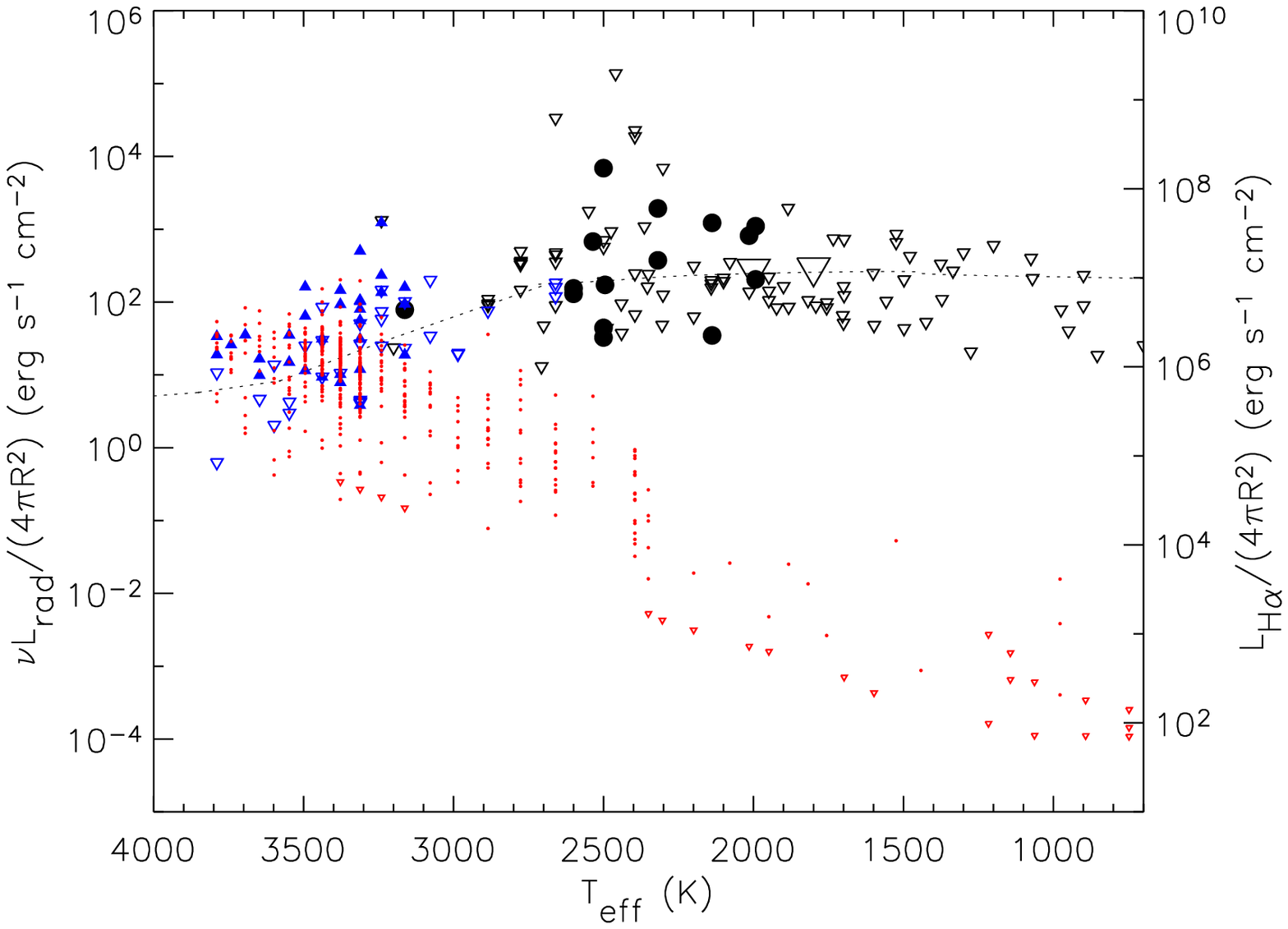}\hfill
   \includegraphics[width=.45\linewidth]{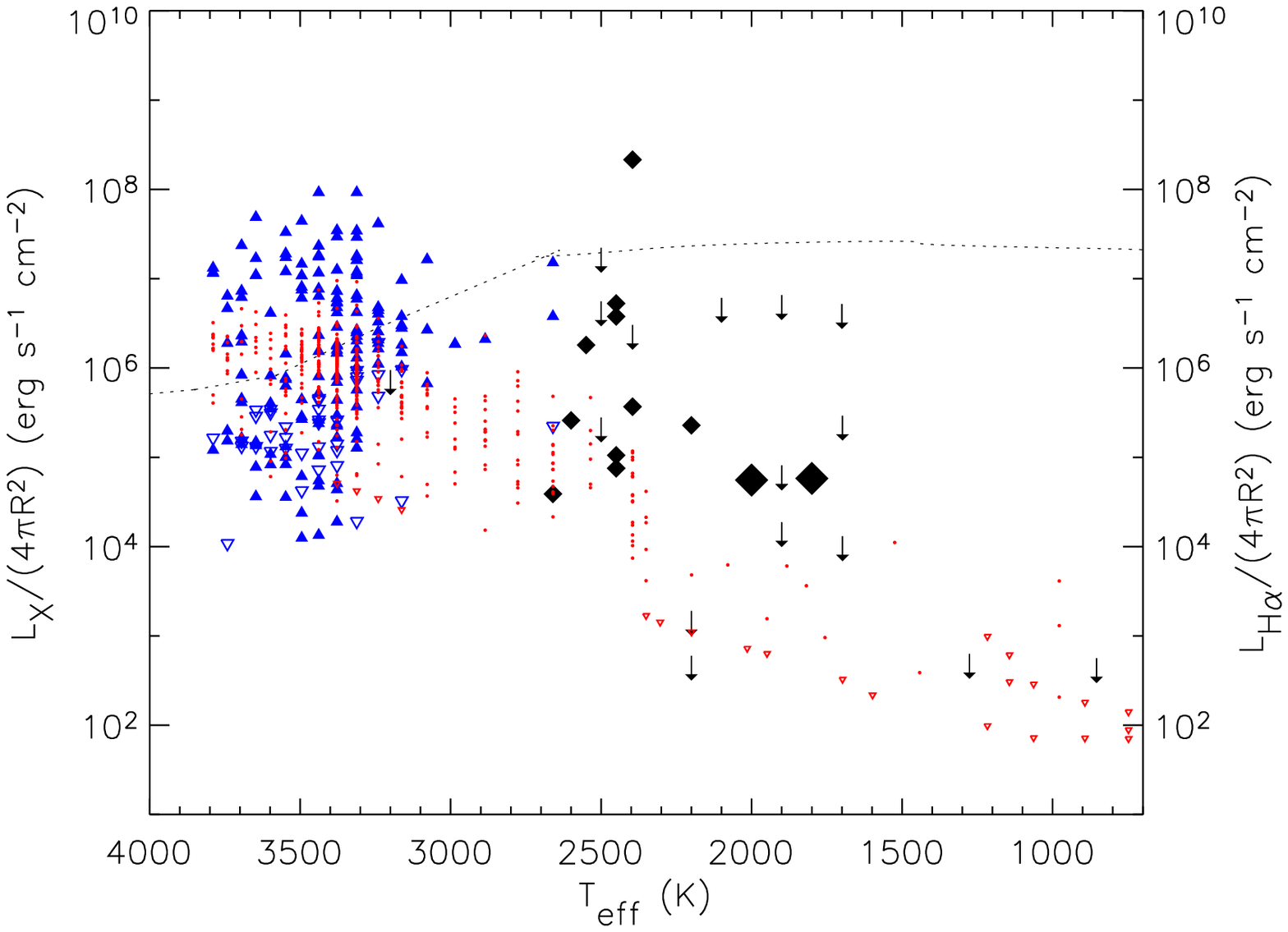}
      \caption{(Top): Luminosity to bolometric luminosity ratio in the radio (left panel) and in the X-rays (right panel). The ratio
      for H$\alpha$ is shown in both panels with its scale on the right $y$-axis and as small symbols in red. 
      (Bottom): Similar as above, for surface fluxes. 
      Ultracool dwarfs 
      are shown in black (filled circles in the radio and diamonds in X-rays for detections, empty downward-pointing triangles 
      for radio upper limits and downward-pointing arrows for X-ray upper limits), 
      while M dwarfs 
      are shown in blue (filled upward-pointing triangles for detections, empty downward-pointing triangles for
      upper limits).
      The Kelu-1AB points (we assumed that the X-ray flux and radio upper limit were either due to Kelu-1A or to Kelu-1B)
      are shown as larger symbols. 
      Note that flaring levels (if known) were purposely discarded.
      The dotted lines show the variations of $L/L_{\rm bol}$ ratios or the surface fluxes vs. $T_{\rm eff}$ for an 
      arbitrary constant luminosity (in radio, X-rays, or H$\alpha$).
      }
         \label{fig:lum}
   \end{figure*}

The top panels of Fig.~\ref{fig:lum} show the $\nu L_{\rm R}/L_{\rm bol}$ and $L_{\rm X}/L_{\rm bol}$ ratios for ultracool dwarfs and
for M dwarfs, and the $L_{\rm H\alpha}/L_{\rm bol}$  ratio for comparison, as a function of effective temperature\footnote{Data taken from the literature; radio:
\citet{krishna99,burgasser05,berger06} and this paper for ultracool dwarfs, \citet{white89,guedel93b,leto00} for M dwarfs; X-rays:
\citet{rutledge00,martin02,fleming03,stelzer03,audard05,berger05,burgasser05,stelzer06a,stelzer06b} 
for ``old'' ultracool dwarfs $\geq 100$~Myr, \citet{doyle89,white89,guedel93b,delfosse98,leto00} for M dwarfs; H$\alpha$: \citet{hawley96,delfosse98,gizis00,burgasser03,mohanty03}. 
If unavailable, $L_{\rm bol}$, $T_{\rm eff}$, and $R/R_\odot$ were calculated from polynomial fits based on some of the above data, or 
from \citet{golimowski04}, and Stefan-Boltzmann's law. 
}.
As noted in previous studies, the X-ray ratio significantly decreases with decreasing effective temperature, in line with the H$\alpha$ ratio. On the other hand, 
the radio ratio increases \citep{berger02,burgasser05}.
However, it should be noted that most radio observations
of L and T dwarfs only provide upper limits \citep[e.g.,][]{berger06}. The top panels also show as a dotted line the $L/L_{\rm bol}$ ratio (in radio, X-rays, or H$\alpha$) 
vs. $T_{\rm eff}$ for an arbitrary constant radio/X-ray/H$\alpha$ luminosity. For ultracool dwarfs, the ratios increase because
their bolometric luminosity depends essentially only on the effective temperature, i.e., $\log L_{\rm bol} \propto \log T_{\rm eff}$.
Indeed, radii in ultracool dwarfs vary little ($R \sim 0.09 R_\odot$).  The $L/L_{\rm bol}$ ratios, therefore, increase
with decreasing $T_{\rm eff}$ for a constant luminosity (in radio, X-rays, or H$\alpha$). The radio upper limits in L and T dwarfs 
are also  consistent with $\nu L_{\rm R}/L_{\rm bol}$ for a constant radio luminosity, suggesting 
that the current radio data do not go deep enough to detect any cut-off, if present. In contrast, the X-ray and H$\alpha$ observations are deep enough to
detect it.

While the $L/L_{\rm bol}$ ratio defines the amount of power radiated in a wavelength regime compared to the bolometric luminosity, and while it is
considered a good measure of magnetic activity in late-type stars, the ratio might be less adequate in ultracool dwarfs. Perhaps the surface flux (i.e., the ratio of the 
luminosity and the dwarf's surface, $L/(4 \pi R^2)$; \citealt{schmitt97}) may better trace magnetic activity at the bottom of the 
main sequence\footnote{ 
Since $R/R_\odot$ is almost constant for ultracool dwarfs, we could replace the surface flux by the luminosity as
an activity indicator. Nevertheless, to allow comparison with early-M dwarfs, which have larger radii, we use the surface flux
in our discussion.}. The bottom panels of Fig.~\ref{fig:lum} show the surface fluxes in radio, X-ray, and H$\alpha$. If the 
luminosity were constant (in radio, X-rays, or H$\alpha$), we would observe no significant decrease. 
This is approximately the case for the radio (despite the lack of detections below 2000~K), in stark contrast with the decrease in H$\alpha$ and X-rays. Note that we again plotted a dotted 
line of the surface flux for an arbitrary  constant luminosity. The slight increase in radio surface flux observed in late-M/early-L dwarfs compared to 
early-M dwarfs is only due to the decrease in radius with decreasing $T_{\rm eff}$ for M dwarfs, but it is consistent with a constant 
luminosity, suggesting that the radio-emitting mechanism is similar in M dwarfs and in detected ultracool dwarfs, and
that it does not lose its emission strength with decreasing $T_{\rm eff}$.
In contrast, the X-ray surface flux and luminosity in ultracool dwarfs declines in a similar fashion as H$\alpha$. 

It appears that the radio emission reaches a plateau in luminosity across M dwarfs and ultracool dwarfs
(at least above 2000~K), while magnetic activity in the chromosphere (H$\alpha$) and in the hot coronal loops (X-rays) declines with decreasing $T_{\rm eff}$.
This result may point toward a different kind of mechanism of radio and X-ray/H$\alpha$ in ultracool dwarfs and in late-type stars, which could explain the observed
deviations of ultracool dwarfs from the G\"udel-Benz $L_{\rm R}-L_{\rm X}$ relation. Cyclotron maser emission is possibly the dominant radio emission mechanism, 
as claimed by \citet{hallinan06,hallinan07}. Such a mechanism was indeed proposed for M flare stars using a dipole magnetic trap model 
\citep{bingham01,kellett02}. The non-detection of Kelu-1 could be due to a lack of sensitivity, or simply
the inclination of Kelu-1's rotation axis does not allow the beam of coherent radio emission to cross our line of sight.
Or perhaps  we observed at too high frequencies (since $\nu_c = 2.8 B$, with $\nu_c$ in GHz and $B$ in kG, 
$\nu_c = 8.4$~GHz requires $B \sim 3$~kG).

\section{Conclusions}
\label{sect:concl}

We have presented the first X-ray detection of an L binary dwarf, Kelu-1AB, while the binary remains undetected in the radio.
The suggested increase in $L_{\rm R}/L_{\rm bol}$ in ultracool dwarfs may be an artifact of the dependence of 
$L_{\rm bol}$ almost solely on $T_{\rm eff}$ in ultracool dwarfs. 
We suggest that the radio luminosity stays constant, at least down to  $T_{\rm eff} \sim 2000$~K, and may drop for cooler temperatures, although current radio surveys of ultracool 
dwarfs lack sensitivity. In contrast, magnetic activity as measured in X-rays and H$\alpha$ slowly declines with decreasing $T_{\rm eff}$. The main
dominant radio emission mechanism in ultracool dwarfs may not be gyrosynchrotron as in earlier-type main-sequence stars but coherent emission by electron 
cyclotron maser.  
The slower spin-down of late-M and L dwarfs
than G-K dwarfs may also lead to a different behavior in the radio and in X-rays for ultracool dwarfs.
There is a need to go deeper in the radio regime (and shorter frequencies if electron-cyclotron maser is the main radio emission mechanism) 
to determine if and when the radio emission declines in the increasingly neutral atmospheres of ultracool L and T dwarfs.

\begin{acknowledgements}
We thank an anonymous referee for useful comments.
M.~A. acknowledges support from a Swiss National Science Foundation Professorship (PP002--110504). 
Support for R.~A.~O. was provided by NASA through Hubble Fellowship grant HF-01189.01 awarded by the STScI, 
which is operated by the Association of Universities for Research in Astronomy, Inc., for NASA, under contract NAS5-26555.
A.~B. and J.~E.~G. acknowledge support by NASA through \cxc\ award GO5-6013. 
The \cxc\ X-ray Observatory Center is operated by the Smithsonian Astrophysical 
Observatory for and on behalf of the NASA under contract NAS8-03060. 
The NRAO is a facility of the National Science Foundation operated under cooperative agreement by Associated Universities, Inc.
The PSI group acknowledges support from the Swiss National Science Foundation (grants 20-58827.99 
and 20-66875.01). M.~A. thanks S.~Wolk, N.~Laslo, B.~Clark, and J.~ Wrobel for their efforts to coordinate the \cxc\ 
and VLA observations.
\end{acknowledgements}

\end{document}